\documentclass[11pt]{amsart}
\usepackage{geometry}                
\usepackage{graphicx}
\usepackage[utf8x]{inputenc}
\usepackage{amssymb}
\usepackage{epstopdf}
\usepackage{listings}
\usepackage{url}
\title{Syntax and Semantics of Babel-17}
\author{Steven Obua}
\date{\today}                                          
\newcommand{\metababel}[1] {\textsl{#1}}
\newcommand{\tabparbox}[1] {\parbox{12cm}{

\vspace{0.1cm}
#1
\vspace{0.1cm}
}}
\newcommand{\patterndescr}[1] {\parbox{9cm}{
\small
\vspace{0.1cm}
#1
\vspace{0.1cm}
}}
\newenvironment{babelcode}[0]{\begin{center}\tt}{\end{center}}

\lstdefinelanguage{babellanguage} 
{morekeywords={lazy, concurrent, force, match, case, exception, begin, end, random, choose,force,true, false, infinity, val, def, this,as,object,if,then,else,elseif,memoize,for,in,yield,while,do,with,nil, div, mod,try,catch,private,import,module,unittest,typedef,typeof,and,or,not, root,native,xor,lens,min,max},
sensitive=true, 
morecomment=[l]{//}, 
morecomment=[s]{/*}{*/}, 
morestring=[b]", 
} 

\lstnewenvironment{babellisting} 
{\small} 
{} 

\newcommand{\babelsrc}[1] {\lstinline!#1!}

\begin{document}


\maketitle

\tableofcontents

\section{Introduction}
The first question that someone who creates a new programming language will  hear from others inevitably is: Why another programming language? Are there not already enough programming languages out there that you can pick from?

I always wanted a programming language 
\begin{itemize}
\item in which I can express myself compactly and without jumping through too many hoops, 
\item that is easy to understand and to learn,
\item in which purely functional programming is the default and not the exception,
\item that supports exceptions,
\item that has pattern matching,
\item that has built-in support for laziness and concurrency,
\item that has built-in support for purely functional lists, vectors, sets and maps,
\item that seamlessly marries structured programming constructs like loops with purely functional programming,
\item that supports objects, modules and data encapsulation,
\item that has mature implementations for all major modern computing platforms and can (and should!) be used for real-world programming,
\item that has a simple mechanized formal semantics,
\item that is beautiful.
\end{itemize}
There is no such language out there. Therefore I decided to create one. When Babel-17 will have reached v1.0 then the above goals will have been achieved. The version described in this document is Babel-17 v0.3.2, so there is still some distance to go. 

Babel-17 is not a radically new or revolutionary programming language. It picks those raisins I like out of languages like Standard ML~\cite{standardml}, Scala~\cite{scala}, Isabelle~\cite{isabelle} (which is not a language, but a system for interactive theorem proving), Alice ML, Java, Javascript, Erlang, Haskell, Lisp, Clojure and Mathematica, and tries to evolve them into a beautiful new design. 

Babel-17 is a functional language. It is not a pure functional language because it contains sources of non-determinism like the ability to pick one of two values randomly. If you take these few sources of non-determinism out of the language, it becomes pure. 

In Babel-17, every value has a type. These types are not checked statically, but only dynamically. 

Babel-17 is also object-oriented in that many aspects of Babel-17 can be explained in terms of sending messages to objects. 

The default evaluation order of Babel-17 is strict. But optional laziness is built into the heart of Babel-17, also, and so is concurrency. 

Babel-17 has pattern matching and exception handling. Both are carefully integrated with each other and with the laziness and concurrency capabilities of Babel-17.

Furthermore, Babel-17 establishes the paradigm of \emph{purely functional structured programming}~\cite{pfsp}  via the notion of linear scope.

In this document  I specify Babel-17 in a mostly informal style: formal notation is sometimes used when it supports clarity. 

Not part of this document, but also part of the overall Babel-17 specification is its formal ANTLR v3.0 grammar; and the (small, but growing) set of unit tests that correspond to the different sections of this document. Both artifacts are available at~\cite{babel17}.

\section{Reference Implementation}

At \cite{babel17} you will find a reference implementation of Babel-17 that fully supports the language as described in this document. There you will also find a Netbeans plugin that has syntax and error-highlighting for Babel-17 programs. As you find your way through this document, it might be helpful to directly try out Babel-17 in Netbeans to see if the interpreter behaves as you would guess from this spec. 
If you find implausible behavior, email your findings to \texttt{bugparade@babel-17.com}. Preferably you would also send along  a unit test that fails because of the observed behavior, but which you think should succeed.

\section{Lexical Matters}
Babel-17 source code is always written UTF-8 encoded. If it ain't UTF-8, it ain't Babel-17. 

Constructors are alphanumeric strings which can also contain underscores, and which must start with a capital letter. 
Identifiers are alphanumeric strings which can also contain underscores, and which must start with a non-capital letter.
The distinction between two constructors or two identifiers is not sensitive to capitalization.

These are the keywords of Babel-17:
\begin{center}
\tt
\begin{tabular}{cccccccc}
begin & end & object & with & if & then & else & elseif \\
while &  for & do & choose & random & yield & match & case \\
as & val & def & in & exception & lazy & concurrent & memoize \\
to & downto & true & false & nil & unittest &  force & this \\
try & catch &  typedef &  typeof &  module &  private &  import & not  \\
and & or & xor & native &  root &  lens &   min & max 
\end{tabular}
\end{center}
Note that keywords are written always in lower case. For example, {\tt bEGIN} is not a legal keyword, but it also isn't an identifier (because of case insensitivity). Also note that {\tt Begin} and {\tt BEGIN} denote the same legal constructor. When we talk about identifiers in the rest of the paper, we always mean identifiers which are not keywords.

You can write down decimal, hexadecimal, binary and octal integers in Babel-17. The following notations all denote the same integer:
\begin{center}
\tt
\begin{tabular}{cccc}
15 & 0xF & 0b1111 &  0o17
\end{tabular}
\end{center}

There is also a decimal syntax for floating point numbers in Babel-17:
\begin{center}
\tt
\begin{tabular}{ccccc}
15.0 & 1.5E1 & 1.5E+1 &  150E-1 & 15e0
\end{tabular}
\end{center}

Strings start and end with a quotation mark. Between the quotation mark, any valid UTF-8 string is allowed that does not contain newline, backslash or quotation mark characters, for example:
\begin{center}
\begin{tabular}{cccc}
\verb+"hello world"+ & \verb+"$1, 2%"+ & \verb+"1 <= 4! <= 5^2"+
\end{tabular}
\end{center}
Newline, backslash and quotation mark characters can be written down in escaped form. Actually, any unicode character can be written down in its escaped form via the hexadecimal value of its codepoint: 
\begin{center}
\begin{tabular}{r|l}
String consisting only of a quotation mark & \verb+"\""+\\\hline
String consisting only of a backslash & \verb+"\\"+ \\\hline
String consisting only of a line feed & \verb+"\n"+ \\\hline
String consisting only of a carriage return & \verb+"\r"+ \\\hline
String consisting only of line feed via 16-bit escaping & \verb+"\u000A"+ \\\hline
String consisting only of line feed via 32-bit escaping & \verb+"\U0000000A"+
\end{tabular}
\end{center}

Multi-line comments in Babel-17 are written between matching pairs of \verb+#(+ and \verb+)#+ and can span several lines. 
Single-line comments start with \verb+##+. 

\emph{Pragmas} start with \verb+#+, just like comments. There are the following pragmas: \verb+#assert+, \verb+#catch+, \verb+#log+, \verb+#print+, \verb+#profile+. 

Finally, here are all ASCII special symbols that can occur in a Babel-17 program:
\newcommand{\mapcurlyleft}{\{|}
\begin{center}
\tt
\begin{tabular}{ccccccccccccc}
\verb+=+ & \verb+==+ & \verb+<>+ & \verb+<+ & \verb+<=+ & \verb+>+& \verb+>=+  &
\verb-+- & \verb+-+ & \verb+*+ & \verb+/+ & \verb+^+ & \verb+;+  \\
\verb-|- & \verb+&+ & \verb+!+ & \verb-++- & \verb+--+& \verb+**+  &
\verb-//- & \verb+,+ & \verb+::+ & \verb+->+ & \verb+=>+& \verb+?+  & \verb+...+  \\
\verb-(- & \verb+)+ & \verb+[+ & \verb+]+ & \verb+{+& \verb+}+  & \verb+.+ &  \verb+:+ &
 \verb+~+ &\verb+:>+ & \verb-+=-&  \verb-++=- & \verb-=+- \\
 \verb-=++- & \verb+*=+ & \verb+**=+ & \verb+=*+& \verb+=**+  & \verb+/=+ &  \verb+//=+ &
 \verb+=/+ &\verb+=//+ &\verb-^=-&  \verb-=^- \\
\end{tabular}
\end{center}
Table~\ref{table:furthersymbols} lists all other Unicode symbols that have a meaning in Babel-17.
\begin{table}
\caption{Further Unicode Symbols}
\begin{tabular}{c|c|c}
\textbf{Unicode Hex Code} & \textbf{Display} & \textbf{ASCII Equivalent}\\\hline
2261 & ≡ & \verb+==+\\
2262 & ≢ & \verb+<>+\\
2264 & ≤ & \verb+<=+\\
2265 &  ≥ & \verb+>=+\\
2237 & :: & \verb+::+\\
2192 & → & \verb+->+\\
21D2 & ⇒ & \verb+=>+\\
2026 & … & \verb+...+\\
\end{tabular}
\label{table:furthersymbols}
\end{table}

\section{Overview of Built-in Types}
Each value in Babel-17 has a unique type. All built-in types are depicted in Table~\ref{table:builtintypes}.
\begin{table}
\caption{Built-in Types of Babel-17}
\begin{tabular}{c|c|c}
\textbf{Name} & \textbf{Type} & \textbf{Description}\\\hline
\metababel {Integer} & \verb+int+ & the type of arbitrary size integer numbers\\
\metababel {Real} & \verb+real+ & the type of real numbers\\
\metababel{Boolean} & \verb+bool+ &  the type consisting of the values \babelsrc{true} and \babelsrc{false}\\
\metababel{String} & \verb+string+ &  the type of valid Unicode strings\\
\metababel{List} & \verb+list+&  the type of lists\\
\metababel{Vector} & \verb+vect+ &  the type of vectors / tuples\\
\metababel{Set} & \verb+set+ &  the type of sets \\
\metababel{Map}& \verb+map+ &  the type of maps \\
\metababel{CExpr} & \verb+cexp+ &  the type of constructed expressions\\
\metababel{Object} & \verb+obj+ &  the type of user-defined objects\\
\metababel{Function} & \verb+fun+ &  the type of  functions\\
\metababel{PersistentException} & \verb+exc+ &  the type of persistent exceptions\\
\metababel{DynamicException} & &  the type of dynamic exceptions\\
\metababel{Type} &  \verb+type+ &  the type of types\\
\metababel{Module} &  \verb+module_+ &  the type of modules\\
\metababel{Lens} &  \verb+lens_+ &  the type of lenses\\
\metababel{Native} &  \verb+native_+ &  the type of native values\\
\end{tabular}
\label{table:builtintypes}
\end{table}
Often we use the name of a type to refer to this type in this specification, but the representation of it in actual Babel-17 is given in the type column. 
The type of a value $v$ can be obtained via 
\begin{babellisting}
typeof $v$
\end{babellisting}

\section{Objects and Functions}
All values in Babel-17 are objects. This means that you can send messages to them. Many values in Babel-17 are functions. This means that you can apply them to an argument, yielding another value. Note that for a value to be an object, it does not need to have type \texttt{obj}, and for a value to be a function,  it does not need to have type \texttt{fun}. 

The result of sending message $m$ to object $v$  is written as
\begin{babellisting}
$v$ . $m$
\end{babellisting}
Here $m$ is an identifier.

The result of applying function $f$ to value $x$ is written as 
\begin{babellisting}
$f$ $x$
\end{babellisting}
Note that $f$ $x$ is equivalent to 
\begin{babellisting}
($f$.apply_) $x$
\end{babellisting}
Therefore any object that responds to the \babelsrc{apply_} message can act as a function.
In the above we could also leave out the brackets because sending messages binds stronger than function application. 

Repeated function application associates to the left, therefore 
\begin{babellisting}
$f$ $x$ $y$ 
\end{babellisting} 
is equivalent to \babelsrc{($f\ x$)\ $y$}.

\section{Exceptions}
Exception handling in Babel-17 mimicks exception handling mechanisms like those which can be found in Java or Standard ML, while adhering at the same time to the functional paradigm.

There are two types of exceptions: \metababel{PersistentException} and  \metababel{DynamicException}.
The difference is that a \metababel{PersistentException}  can be treated as part of any data structure, and is passed around just like any other value, while a \metababel{DynamicException} can never be part of a data structure and has special evaluation semantics. 

Exceptions in Babel-17 are not that special at all but mostly just another type of value.  Let us write \metababel{PersistentException $v$} for a \metababel{PersistentException} with parameter $v$,
and  \metababel{DynamicException $v$} for a \metababel{DynamicException} with parameter $v$. 
We also write \metababel{Exception $v$} for an exception with parameter $v$. The parameter $v$ is a \emph{non-exceptional} value; this means that $v$ can be any Babel-17 value except one of type \metababel{DynamicException}. Note that a value of type \metababel{PersistentException} is therefore non-exceptional.

The following deterministic rules govern how exceptions behave with respect to sending of messages, function application, laziness and concurrency:
\begin{babelcode}
\begin{tabular}{rcl}
(\metababel{Exception $v$}).$m$ & $\leadsto$ & \metababel{DynamicException $v$}\\
(\textsl{DynamicException} $v$) $x$ & $\leadsto$ & \metababel{DynamicException $v$}\\
$f$ (\metababel{DynamicException $v$}) & $\leadsto$ & \metababel{DynamicException $v$}  {\rm where $f$ is non-exceptional}\\
(\metababel{PersistentException $v$}) $g$ & $\leadsto$ & \metababel{DynamicException $v$}  {\rm where $g$ is non-exceptional}\\
exception $v$ & $\leadsto$ & \metababel{DynamicException $v$}\\
lazy (\metababel{Exception $v$}) & $\leadsto$ & \metababel{PersistentException $v$}\\
concurrent (\metababel{Exception $v$}) & $\leadsto$ & \metababel{PersistentException $v$}\\
force $v$  & $\leadsto$ & $v$  {\rm for all $v$, including exceptions}
\end{tabular}\\
\end{babelcode}
Exceptions in Babel-17 are created with the expression \babelsrc{exception $\ v$}. Creating exceptions in Babel-17 corresponds to \emph{raising} or \emph{throwing} exceptions in other languages. Catching an exception can be done via \babelsrc{match} or via \babelsrc{try-catch}. Both constructs will be described later.

In the next section, we will describe the  \babelsrc{lazy} and \babelsrc{concurrent} expressions of Babel-17. They are the reason why exceptions are divided into dynamic and persistent ones. 

\section{Laziness and Concurrency}
The default evaluation mechanism of Babel-17 is strict. This basically means that the arguments of a function are evaluated before the function is applied to them. Babel-17 has two constructs to change this default behaviour. 

The expression
\begin{babellisting}
lazy $e$
\end{babellisting}
is not evaluated until it is actually needed. When it is needed, it evaluates to whatever $e$ evaluates to, with the exception of dynamic exceptions, which are converted to persistent ones.  

The expression
\begin{babellisting}
concurrent $e$
\end{babellisting}
also evaluates to whatever $e$ evaluates to, again with the exception of dynamic exceptions which are converted to persistent ones. This evaluation will happen concurrently.  

One could think that apart from obvious different performance characteristics, the expressions \babelsrc{lazy $\ e$} and \babelsrc{concurrent $\ e$} are equivalent. This is not so.
If $e$ is a non-terminating expression then, even if  the value of  \babelsrc{concurrent  $\ e$} is never needed during program execution, it might still lead to a non-terminating program execution. In other words, the behaviour of \babelsrc{concurrent $\ e$}  is unspecified for non-terminating $e$.

Sometimes you want to explicitly force the evaluation of an expression. In those situations you use the expression
\begin{babellisting}
force $e$
\end{babellisting}
which evaluates to $e$. So semantically, \babelsrc{force} is just the identity function. 

We mentioned before that lazy and concurrent expressions are the reason why exceptions are divided into dynamic and persistent ones. To motivate this, look at the expression
\begin{babellisting}
fst (0, lazy (1 div 0))
\end{babellisting}
Here the function \babelsrc{fst} is supposed to be a function that takes a pair and returns the first element of this pair. So what would above expression evaluate to? Obviously, \babelsrc{fst} does not need to know the value of the second element of the pair as it depends only on the first element, so above expression evaluates just to 0. Now, if \babelsrc{lazy} was semantically just the identity function, then we would have
\begin{babellisting}
0 = fst (0, lazy (1 div 0)) = fst (0, 1 div 0) = fst (0, exception DomainError) 
   = fst (exception DomainError) = exception DomainError
\end{babellisting}
Obviously, $0$ should not be the same as an exception, and therefore \babelsrc{lazy} cannot be the identity function, but converts dynamic exceptions into persistent ones. For a dynamic exception $e$ the equation
\begin{babellisting}
(0, $e$) = $e$
\end{babellisting}
holds. For a persistent exception $e$ this equation does not hold, and therefore the above chain of equalities is broken.

\section{Lists and Vectors}
For $n \ge 0$, the expression
\begin{babellisting}
[$e_1$, $\ldots$, $e_n$]
\end{babellisting}
denotes a \emph{list} of $n$ elements. 

The expression
\begin{babellisting}
($e_1$, $\ldots$, $e_n$)
\end{babellisting}
denotes a  \emph{vector} of $n$ elements, at least for $n \neq 1$. 
For $n=1$, there is a problem with notation, though, because \babelsrc{($e$)}  is equivalent to $e$. Therefore there is the special notation \babelsrc{($e$,)} for vectors which consist of only one element.

The difference between lists and vectors is that they have different performance characteristics for the possible operations on them. Lists behave like simply linked lists, and vectors behave like arrays. Note that all data structures in Babel-17 are immutable. 

Another way of writing down lists is via the right-associative \babelsrc{::} constructor:
\begin{babellisting}
$h$::$t$
\end{babellisting}
Here $h$ denotes the \emph{head} of the list and $t$ its \emph{tail}. Actually, note that the expression \babelsrc{[$e_1$, $\ldots$, $e_n$]} is just syntactic sugar for the expression
\begin{babellisting}
$e_1$::$e_2$::$\ldots$::$e_n$::[]
\end{babellisting}
Dynamic exceptions cannot be part of a list but are propagated:
\begin{babelcode}
\begin{tabular}{rcll}
(\metababel{DynamicException $v$})::$t$ & $\leadsto$ & \metababel{DynamicException $v$}&\\
$h$::(\metababel{DynamicException $v$}) & $\leadsto$ & \metababel{DynamicException $v$}& {\rm where $h$ is non-exceptional}\\
\end{tabular}\\
\end{babelcode}
Note that when the tail $t$ is not a list, we identify \babelsrc{$h$::$t$} with  \babelsrc{$h$::$t$::[]} . 

\section{CExprs}
A \metababel{CExpr} is a constructor $c$ together with a parameter $p$, written $c$ $p$. It is allowed to leave out the parameter $p$, which then defaults to \babelsrc{nil}. For example,  \babelsrc{HELLO} is equivalent to \babelsrc{HELLO} \babelsrc{nil}. A constructor $c$ cannot have a dynamic exception as its parameter, therefore we have:
\begin{babelcode}
\begin{tabular}{rcll}
$c$ (\metababel{DynamicException $v$}) & $\leadsto$ & \metababel{DynamicException $v$}&\\
\end{tabular}\\
\end{babelcode}

\section{Pattern Matching}
Maybe the most powerful tool in Babel-17 is pattern matching. You use it in several places, most prominently in the \babelsrc{match} expression which has the following syntax:
\begin{babellisting}
match $e$ 
  case $p_1$ => $b_1$
  $\vdots$
  case $p_n$ => $b_n$
end  
\end{babellisting}
Given a value $e$, Babel-17 tries to match it to the patterns $p_1$, $p_2$, $\ldots$ and so on sequentially in that order. If $p_i$ is the first pattern to match, then the result of \babelsrc{match} 
is given by the block expression $b_i$. If none of the pattern matches then there are two possible outcomes:
\begin{enumerate}
\item If $e$ is a dynamic exception, then the value of the match is just $e$.
\item Otherwise the result is a dynamic exception with parameter \babelsrc{NoMatch}.
\end{enumerate}
A few of the pattern constructs  incorporate arbitrary value expressions. When these expressions raise exceptions, they are propagated up.

So what does a pattern look like? Table~\ref{table:basicpatterns}  and  Table~\ref{table:collectionpatterns} list all ways of building a pattern.
\begin{table}
\caption{General Patterns}
\begin{tabular}{c|l}
\textbf{Syntax} & \textbf{Description}\\\hline
\babelsrc{_} & \patterndescr{the underscore symbol matches anything but a dynamic exception}\\\hline
\babelsrc{$x$} & \patterndescr{an identifier $x$ matches anything but a dynamic exception and binds the matched expression to $x$}\\\hline
\babelsrc{($x\ $ as $\ p$)} & \patterndescr{matches $p$, and binds the successfully matched value to $x$; the match fails if $p$ does not match or if the matched value is a dynamic exception}\\\hline
\babelsrc{$z$} & \patterndescr{an integer number $z$, like \babelsrc{0} or \babelsrc{42} or \babelsrc{-10}, matches just that number $z$}\\\hline
\babelsrc{$c\ $ $p$} & \patterndescr{matches a \metababel{CExpr} with constructor $c$ if the parameter of the \metababel{CExpr} matches $p$; instead of \babelsrc{$c\ $ $\_$} you can just write \babelsrc{$c$}}\\\hline
\babelsrc{$s$} & \patterndescr{a string $s$, like \babelsrc{"hello"}, matches just that string $s$}\\\hline
\babelsrc{($p$)} & \patterndescr{same as $p$} \\\hline
\babelsrc{($p\ $ if $\ e$)} & \patterndescr{matches any non-exceptional value that matches $p$, but only if $e$ evaluates to \babelsrc{true}; identifiers bound in $p$ can be used in $e$}\\\hline
\babelsrc{(val $\ e$)} & \patterndescr{matches any non-exceptional  value which is equivalent to $e$; in case an \texttt{Unrelated}-exception is thrown, this pattern just does not match, and does \emph{not} propagate up the exception}\\\hline
\babelsrc{($c\ $ \! $\ p$)} & \patterndescr{let $v$ be the value to be matched; then the match succeeds if the result $v$.\babelsrc{destruct\_}  $c$ matches $p$}\\\hline
\babelsrc{($c\ $ \!)} & \patterndescr{short for \babelsrc{($c\ $ \! \_)} }\\\hline
\babelsrc{($f\ $ ? $\ p$)} & \patterndescr{$f$ is applied to the value to be matched; the match succeeds if the result of the application matches $p$}\\\hline
\babelsrc{($f\ $ ?)} & \patterndescr{short for \babelsrc{($f\ $ ? true)} }\\\hline
\babelsrc{\{$m_1\ $ = $\ p_1$,$\ \ldots\ $,$\ m_n\ $ = $\ p_n$\}} & \patterndescr{matches a value of type \texttt{obj} that has exactly the messages $m_1$,.., $m_n$ such that the message values match the given patterns}\\\hline
\babelsrc{\{$m_1\ $ = $\ p_1$,$\ \ldots\ $,$\ m_n\ $ = $\ p_n$,  $\ \delta$\}} & \patterndescr{matches a value of type \texttt{obj} that has the messages $m_1$,.., $m_n$ such that the message values match the given patterns}\\\hline
\babelsrc{nil} & \patterndescr{matches the empty object}\\\hline
\babelsrc{exception} $p$ & \patterndescr{matches any exception such that its parameter matches $p$}\\\hline
\babelsrc{$(p : t)$} & \patterndescr{matches anything that has (or is auto-convertible to) type $t$ and matches $p$}\\\hline
\babelsrc{$(p : (e))$} & \patterndescr{matches anything that matches $p$ and is or auto-converts to the type that $e$ evaluates to}\\\hline
\babelsrc{$(t\ p)$} & \patterndescr{inner-value pattern; see Section~\ref{sec:typedefs}}
\end{tabular}
\label{table:basicpatterns}
\end{table}
\begin{table}
\caption{Collection Patterns}
\begin{tabular}{c|l}
\textbf{Syntax} & \textbf{Description}\\\hline
\babelsrc{[$p_1$, $\ \ldots\ $, $\ p_n$]} &  \patterndescr{matches a list consisting of $n \ge 0$ elements, such that element $e_i$ of the list is matched by pattern $p_i$}\\\hline
\babelsrc{($p_1$, $\ \ldots\ $, $\ p_n$)} &  \patterndescr{matches a vector consisting of $n=0$ or $n \ge 2$ elements, such that element $e_i$ of the list is matched by pattern $p_i$}\\\hline
\babelsrc{[$p_1$, $\ \ldots\ $, $\ p_n$, $\ \delta$]} &  \patterndescr{matches a list consisting of at least $n \ge 1$ elements, such that the first $n$ elements $e_i$ of the list are matched by the patterns $p_i$}\\\hline
\babelsrc{($p_1$, $\ \ldots\ $, $\ p_n$, $\ \delta$)} &  \patterndescr{matches a vector consisting of at least $n \ge 1$ elements, such that the first $n$ elements $e_i$ of the vector are matched by the patterns $p_i$}\\\hline
\babelsrc{($p$, )} &  \patterndescr{matches a vector consisting of a single element that matches the pattern $p$.} \\\hline
\babelsrc{($h$::$t$)} & \patterndescr{matches a non-empty list such that $h$ matches the head of the list and $t$ its tail}\\\hline
\babelsrc{\{$p_1$,$\ \ldots\ $,$\ p_n$\}} & \patterndescr{see section~\ref{sec:setsandmaps}} \\\hline
\babelsrc{\{$p_1$,$\ \ldots\ $,$\ p_n\ $, $\ \delta$\}} & \patterndescr{see section~\ref{sec:setsandmaps}} \\\hline
\babelsrc{\{$q_1\ $ -> $\ p_1$,$\ \ldots\ $,$\ q_n\ $ -> $\ p_n$\}} & \patterndescr{see section~\ref{sec:setsandmaps}}\\\hline
\babelsrc{\{$q_1\ $ -> $\ p_1$,$\ \ldots\ $,$\ q_n\ $ -> $\ p_n$, $\ \delta$\}} & \patterndescr{see section~\ref{sec:setsandmaps}}\\\hline
\babelsrc{\{->\}} & \patterndescr{see section~\ref{sec:setsandmaps}}\\\hline
\babelsrc{\(for $\ p_1$,$\ \ldots\ $, $\ p_n\ $end\)} & \patterndescr{see section~\ref{sec:loops}}\\\hline
\babelsrc{\(for $\ p_1$,$\ \ldots\ $, $\ p_n$, $\ \delta\ $end\)} & \patterndescr{see section~\ref{sec:loops}}
\end{tabular}
\label{table:collectionpatterns}
\end{table}

In this table of pattern constructions we use the $\delta$-pattern $\delta$. This pattern stands for "the rest of the entity under consideration" and can be constructed by the following rules:
\begin{enumerate}
\item The ellipsis \babelsrc{...} is a $\delta$-pattern that matches any rest.
\item If $\delta$ is a $\delta$-pattern, and $x$ an identifier, then \babelsrc{($x\ $ as $\ \delta$)} is a $\delta$-pattern.
\item If $\delta$ is a $\delta$-pattern, and $e$ an expression, then \babelsrc{($\delta\ $ if $\ e$)} is a $\delta$-pattern.
\end{enumerate}

Note that pattern matching does not distinguish between vectors and lists. A pattern that looks like a vector can match a list, and vice versa.

Besides the \babelsrc{match} construct, there is also the \babelsrc{try} construct. While \babelsrc{match} can handle exceptions, most of the time  it is more convenient to use \babelsrc{try} for this purpose. The syntax is 
\begin{babellisting}
try 
  $s_1$
  $\ldots$
  $s_m$
catch 
  case $p_1$ => $b_1$
  $\vdots$
  case $p_n$ => $b_n$
end  
\end{babellisting}
The meaning of the above is similar to the meaning of
\begin{babellisting}
match 
  begin 
    $s_1$
    $\ldots$
    $s_m$
  end
case (exception $p_1$) => $b_1$
  $\vdots$
case (exception $p_n$) => $b_n$
case x => x
end  
\end{babellisting}
except for two differences: 
\begin{enumerate}
\item  the latter expression might not be legal Babel-17 because of linear scoping violations,
\item \babelsrc{try} does not catch persistent exceptions.
\end{enumerate}

\section{Non-Determinism in Babel-17}

One source of non-determinism in Babel-17 is \emph{probabilistic} non-determism. The expression
\begin{babellisting}
random $n$
\end{babellisting}
returns for an integer $n > 0$ a number between $0$ and $n-1$ such that each number between $0$ and $n-1$ has equal chance of being returned. If $n$ is a dynamic exception, then this exception is propagated, if $n$ is a non-exceptional value that is not an integer $> 0$ then the result is an exception with parameter \babelsrc{DomainError}.

Another source of non-determinism is choice. The expression
\begin{babellisting}
choose $l$
\end{babellisting}
takes a non-empty collection $l$ and returns a member of the collection. The choice operator makes it possible to optimize evaluation in certain cases. For example, in 
\begin{babellisting}
choose (concurrent a, concurrent b)
\end{babellisting}
the evaluator could choose the member that evaluates quicker.

There is a third source of non-determinism which has its roots in the module loading mechanism (see Section~\ref{modules}). 

Babel-17 has been designed such that if you take the above sources of non-determinism out of the language, it becomes purely functional. If in your semantics of Babel-17 you replace the concept of value by the concept of a probability distribution over values and a set of values, you might be able to view Babel-17 as a purely functional language even \emph{including} \babelsrc{random} and \babelsrc{choose}.

\section{Block Expressions}
So far we have only looked at expressions. We briefly mentioned the term \emph{block expressions} in the description of the \babelsrc{match} function, though.  We will now introduce and explain block expressions.

Block expressions can be used in several places as defined by the Babel-17 grammar. For example, they can be used in a \babelsrc{match} expression to define the value that corresponds to a certain case. But block expressions can really be used just everywhere where a normal expression is allowed: 
\begin{babellisting}
begin
  $b$
end
\end{babellisting}
is a normal expression where $b$ is a block expression. A block expression has the form
\begin{babellisting}
$s_1$
$\vdots$
$s_n$
\end{babellisting}
where the $s_i$ are \emph{statements}. In a block expression both newlines and semicolons can be used to separate the statements from each other.

Statements are Babel-17's primary tool for introducing identifiers. There are several kinds of statements. Three of them will be introduced in this section.

First, there is the \babelsrc{val}-statement which has the following syntax:
\begin{babellisting}
val $p$ = $e$
\end{babellisting}
Here $p$ is a pattern and $e$ is an expression. Its meaning is that first $e$ gets evaluated. If this results in a dynamic exception, then the result of the block expression that the \babelsrc{val}-statement is part of will be that dynamic exception. Otherwise, the result of evaluating $e$ is matched to $p$. If the match is successful then all identifiers bound by the match can be used in later statements of the block expression. If the match fails, then the value of the containing block expression is the dynamic exception \babelsrc{NoMatch}.

Second, there is the \babelsrc{def}-statement which obeys the following syntax for defining the identifier $id$:
\begin{babellisting}
def $id$ $arg$ = $e$
\end{babellisting}
The $arg$ part is optional. If $arg$ is present, then it must be a pattern. Let us call those definitions where $arg$ is present a \emph{function definition}, and those definitions where $arg$ is not present a \emph{simple definition}.  

Per block expression and identifier $id$ there can be either a single simple definition, or there can be several function definitions. If there are multiple function definitions for the same identifier in one block expression, then they are bundled in that order to form a single function.

The defining expressions in  \babelsrc{def}-statements can recursively refer to the other identifiers defined by  \babelsrc{def}-statements in the same block expression. This is the main difference between definitions via \babelsrc{def} and definitions via \babelsrc{val}. Only those \babelsrc{val}-identifiers are in scope that have been defined \emph{previously}, but \babelsrc{def}-identifiers are in scope throughout the whole block expression. Table~\ref{table:legaldef} exemplifies this rule.
\begin{table}
\caption{Legal and illegal definitions}
\begin{tabular}{c@{\hspace{1cm}}c@{\hspace{1cm}}c@{\hspace{1cm}}c}
\begin{babellisting}
val x = y       
val y = 0
\end{babellisting} &
\begin{babellisting}
def x = y       
val y = 0
\end{babellisting} &
\begin{babellisting}
val x = y       
def y = 0
\end{babellisting} &
\begin{babellisting}
def x = y       
def y = 0
\end{babellisting} \\[0.5cm]
\emph{illegal} &
\emph{illegal} &
\emph{legal} &
\emph{legal} 
\end{tabular}
\label{table:legaldef}
\end{table}

Let us assume that  a block expression contains multiple function definitions for the same identifier \babelsrc{f}:
\begin{babellisting}
def f $p_1$ = $e_1$ 
   $\vdots$
def f $p_n$ = $e_n$ 
\end{babellisting}
Then this is (almost) equivalent to 
\begin{babellisting}
def f x = 
  match x
    case $p_1$ => $e_1$
      $\vdots$
    case $p_n$ => $e_n$ 
  end
\end{babellisting}
where \babelsrc{x} is fresh for the $p_i$ and $e_i$. The slight difference between the two notations is that arguments that match none of the patterns will result in a \babelsrc{DomainError} exception for the first notation, and in a \babelsrc{NoMatch} exception for the second.

While definitions only define a single entity per block and identifier, it is OK to have multiple \babelsrc{val}-statements for the same identifier in one block expression, for example like that:
\begin{babellisting}
val x = 1
val x = (x, x)
x
\end{babellisting}
The above block expression evaluates to \babelsrc{(1,1)}; later \babelsrc{val}-definitions overshadow earlier ones.
But note that neither
\begin{babellisting}
val x = 1
def x = 1
\end{babellisting}
nor
\begin{babellisting}
def x = 1
val x = 1
\end{babellisting}
are legal.

Another difference between \babelsrc{val} and \babelsrc{def} is observed by the effects of non-determinism:
\begin{babellisting}
val x = random 2
(x, x)
\end{babellisting}
will evaluate either to \babelsrc{(0, 0)} or to \babelsrc{(1, 1)}. But 
\begin{babellisting}
def x = random 2
(x, x)
\end{babellisting}
can additionally also evaluate to \babelsrc{(0, 1)} or to \babelsrc{(1, 0)} because $x$ is evaluated each time it is used.

Let us conclude this section by describing the third kind of statement. It has the form
\begin{babellisting}
yield $e$
\end{babellisting}
where $e$ is an expression. There is also an abbreviated form of the \babelsrc{yield}-statement which we have already used several times in this section:
\begin{babellisting}
$e$
\end{babellisting}
The semantics of the \babelsrc{yield}-statement is that it appends a further value to the value of the block expression it is contained in. Block expressions have a value just like all other expressions in Babel-17. It is obtained by "concatenating" all values of the \babelsrc{yield}-statements in a block expression. The value of 
\begin{babellisting}
begin
end
\end{babellisting}
is the empty vector \babelsrc{()}. The value of 
\begin{babellisting}
begin
  yield $a$
end
\end{babellisting}
is $a$. The value of 
\begin{babellisting}
begin
  yield $a$
  yield $b$
end
\end{babellisting}
is the vector \babelsrc{($a$,$\ b$)} of length 2 and so on. 

In this section we have introduced the notions of block expressions and statements. Their full power will be revealed in later sections of this document.

\section{Anonymous Functions}
So far we have seen how to define named functions in Babel-17. Sometimes we do not need a name for a certain function, for example when the code that implements this function is actually just as easy to understand as any name for the function. We already have the tools for writing such nameless, or anonymous, functions:
\begin{babellisting}
begin
  def sqr x = x * x 
  sqr
end
\end{babellisting}
is an expression denoting the function that squares its argument. There is a shorter and equivalent way of writing down the above:
\begin{babellisting}
x => x * x
\end{babellisting}
In general, the syntax is
\begin{babellisting}
$p$ => $e$
\end{babellisting}
where $p$ is a pattern and $e$ an expression. The above is equivalent to
\begin{babellisting}
begin
  def f $p$ = $e$ 
  f
end
\end{babellisting}
where \babelsrc{f} is fresh for $p$ and $e$.

There is also a syntax for anonymous functions which allows for several cases:
\begin{babellisting}
(case $p_1$ => $b_1$
    $\vdots$
 case $p_n$ => $b_n$)
\end{babellisting}
is equivalent to 
\begin{babellisting}
begin
  def f $p_1$ = begin $b_1$ end
    $\vdots$
  def f $p_n$ = begin $b_n$ end
  f
end
\end{babellisting}
where $\babelsrc{f}$ is fresh for the $p_i$ and $b_i$.

\section{Object Expressions}
\emph{Object expressions} have the following syntax:
\begin{babellisting}
object 
  $s_1$
  $\vdots$
  $s_n$
end
\end{babellisting}
The $s_i$ are statements.  Those $s_i$ that are  \babelsrc{def} statements define the set of messages that the object responds to.
Often you do not want to create objects from scratch but by modifying other already existing objects:
\begin{babellisting}
object + parents
  $s_1$
  $\vdots$
  $s_n$
end
\end{babellisting}
The expression \babelsrc{parents} must evaluate to either a list, a vector or a set. The members of  \babelsrc{parents}
are considered to be the parents of the newly created object, in the order induced by the collection. The idea is that the created object not only understands the messages defined by the $s_i$, but also the messages of the parents. Messages defined via $s_i$ shadow messages of the parents. The messages of an earlier parent shadow the messages of a later one. 

The keyword \babelsrc{this} can be used only in \babelsrc{def}-statements of an object and points to the object that the original message has been sent too.

\noindent There is also a way to denote record-like objects:
\begin{babellisting}
{ $m_1$ = $v_1$, $\ldots$, $m_n$ = $v_n$ }
\end{babellisting}
This is equivalent to:
\begin{babellisting}
begin
  val ($w_1$, $\ldots$, $w_n$) = ($v_1$, $\ldots$, $v_n$)
  object
    def $m_1$ = $w_1$
    $\cdots$
    def $m_n$ = $w_n$
  end
end
\end{babellisting}
The empty object is denoted by
\begin{babellisting}
nil
\end{babellisting}
which is equivalent to
\begin{babellisting}
object end
\end{babellisting}

\section{Boolean Operators}

Babel-17 provides the usual boolean operators. They are just syntactic sugar for certain \babelsrc{match} expressions; the exact translations are given in Table~\ref{tab:booleanops}. Furthermore, there is the \babelsrc{xor} operator which is defined in the usual way.
\begin{table}
\caption{Boolean Operators}
\begin{tabular}{c|c|c}
\babelsrc{not a} & \babelsrc{a and b} &\babelsrc{a or b} \\\hline
 \small
\begin{babellisting}
match a 
  case true => 
    false 
  case false => 
    true
  case _ => 
    exception DomainError
end
\end{babellisting}
 & 
 \small
\begin{babellisting}
match a 
  case true => 
    match b 
      case true => 
        true
      case false => 
        false
      case _ => 
        exception DomainError
    end 
  case false => 
    false
  case _ => 
    exception DomainError
end
\end{babellisting}
 & 
 \small
\begin{babellisting}
match a 
  case false => 
    match b 
      case true => 
        true
      case false => 
        false
      case _ => 
        exception DomainError
    end 
  case true => 
    true
  case _ => 
    exception DomainError
end
\end{babellisting}
\end{tabular}
\label{tab:booleanops}
\end{table}
Babel-17 also has \babelsrc{if}-expressions with the following syntax:
\begin{babellisting}
if $b_1$ then
  $e_1$
elseif $b_2$ then
  $e_2$
  $\vdots$
elseif $b_n$ then
  $e_n$
else
  $e_{n+1}$
end
\end{babellisting}
The \babelsrc{elseif}-branches are  optional. They can be eliminated in the obvious manner via nesting, so that we only need to give the semantics for the expression 
\begin{babellisting}
if $b$ then $e_1$ else $e_2$ end
\end{babellisting}
The meaning of above expression is defined to be
\begin{babellisting}
match $b$ 
  case true => $e_1$
  case false => $e_2$
  case _ => exception DomainError 
end
\end{babellisting}
Actually, the  \babelsrc{else} branch is also optional. The notation
\begin{babellisting}
if $b$ then $e$  end
\end{babellisting}
is shorthand for
\begin{babellisting}
if $b$ then $e$  else end
\end{babellisting}

\section{Order}\label{sec:order}
Babel-17 has a built-in partial order which is defined in terms of the operator $\sim$. The expression $a \sim b$ returns an \metababel{Integer}; the usual relational  operators are defined in Table~\ref{tab:relops}.
\begin{table}
\caption{Relational Operators}
\begin{tabular}{c|c}
\textbf{Syntax} & \textbf{Semantics} \\\hline
$a < b$ & $(a \sim b) < 0$\\
$a == b$ & $(a \sim b) == 0$\\
$a > b$ & $(a \sim b) > 0$ \\
$a <= b$ &  $(a \sim b) <= 0$ \\
$a >= b$ & $(a \sim b) >= 0$ \\
$a$ $<>$ $b$ & $(a \sim b)$ $<>$ $0$ \\
\end{tabular}
\label{tab:relops}
\end{table}

It is possible to chain relational operators like this:
\begin{babellisting}
$a$ <= $b$ <= $c$ > $d$ <> $e$ 
\end{babellisting}
Intuitively, the above means 
\begin{babellisting}
$a$ <= $b$ & $b$ <= $c$  & $c$ > $d$  & $d$ <> $e$ .
\end{babellisting}
Note that  we always evaluate the operands  of relational operators, even chained ones, only once. For example, the precise semantics of $a <= b <= c <= d <= e$ is 
\begin{babellisting}
begin
  val t = $a$
  val u = $b$
  t <= u &
  begin
    val v = $c$
    u <= v & 
    begin
      val w = $d$
      v <= w & w <= $e$
    end
  end
end
\end{babellisting}
In the above, \babelsrc{t}, \babelsrc{u}, \babelsrc{v} and \babelsrc{w} are supposed to be fresh identifiers.  Also note that if there are operands that are dynamic exceptions, then the result of a comparison is a dynamic exception with the same parameter as the first such operand (from left to right).

It is possible that two values $a$ and $b$ are not related with respect to the built-in order. In this case, $a \sim b$ throws an \emph{Unrelated}-exception. Taking this into account, the definition of $a$ $op$ $b$ for $op$ $\in \{<, >, <=, >=\}$ is
\begin{babellisting}
match a $\sim$ b 
  case (u if u $op$ 0) => true
  case _ => false
end
\end{babellisting}
The cases $op$ $\in \{==, <>\}$ do not propagate the \emph{Unrelated}-exception, and are defined like this:
\begin{babellisting}
match a $\sim$ b 
  case (u if u $op$ 0) => true
  case _ => false
  case (exception Unrelated) => true $op$ false
end
\end{babellisting}

In principle, two values $a$ and $b$ are unrelated if they have different types. This statement is weakened by the existence of automatic type conversion (Section~\ref{sec:automatictypeconversion}): if $a$ and $b$ have different types, it is checked if $a$ can be automatically converted to something of type \babelsrc{typeof\ $b$}, or if $b$ can be automatically converted into something of type \babelsrc{typeof\ $a$}. If so, then one of these conversions is done and the result is compared with the other value.

The built-in partial order has the following properties:
\begin{itemize}
\item if $a$ and $b$ have the same type, and this type is \texttt{obj} or user-defined (see Section ~\ref{sec:typedefs} for how to define your own types), then $a \sim b$ is equivalent to
\begin{babellisting}
    match a.compare_ b
      case u : int => u
      case _ => exception Unrelated
      case (exception _) => exception Unrelated
    end
\end{babellisting}
if $a$ has a \texttt{compare\_} message; if not then $a$ and $b$ are compared according to the number, names and contents of their messages
\item persistent exceptions are partially ordered by their parameter 
\item booleans are totally ordered and \babelsrc{false} $<$ \babelsrc{true} holds
\item integers are totally ordered in the obvious way
\item reals are partially ordered as described in Section~\ref{sec:reals}
\item strings are totally ordered via lexicographic ordering
\item lists are partially ordered by the lexicographic ordering 
\item vectors are partially ordered by the lexicographic ordering 
\item constructed expressions are partially ordered by representing them by the pair consisting of constructor and parameter
\item $f \neq g$ for all values $f$ and $g$ of type \texttt{fun}
\item sets are partially ordered by first comparing their sizes, and then their elements 
\item maps are partially ordered by first comparing their sizes, then their keys, and then their corresponding values
\item types are totally ordered by their names
\item modules are totally ordered by their names
\end{itemize}

There are built-in operators \babelsrc{min} and \babelsrc{max} that compute the minimum and the maximum of a \emph{collection} (Section~\ref{sec:loops} defines what a collection is). For example,
\begin{babellisting}
  min (1, 2) == max (-1, 1, 0) == 1
\end{babellisting}

\section{Sets and Maps}\label{sec:setsandmaps}
Sets and maps are built into Babel-17. For example, the set consisting of 3, 42 and 15 can be written as
\begin{babellisting}
{42, 15, 3}
\end{babellisting}
Sets are always sorted. The sorting order is Babel-17's built-in partial order, and every set forms a totally ordered subdomain of this partial order. Future versions of Babel-17 might allow the set order to be different from the built-in partial order. Adding or removing an element $e$ from a set $S$ is only well-defined when the elements of $S$ together with $e$ are totally ordered by the partial order of $S$. The same holds for testing if an element is in a set. 

Maps map finitely many keys to values. For example, the map that maps 1 to 2 and 4 to 0 is written as
\begin{babellisting}
{1 -> 2, 4 -> 0}
\end{babellisting}
The empty map is denoted by 
\begin{babellisting}
{->}
\end{babellisting}
Maps also have always a partial order associated with them, in the current version of Babel-17 this is always Babel-17's built-in partial order.  Operations on maps are only well-defined if all keys of the map together with all other involved keys are totally ordered by the associated order. 

Pattern matching is available also for sets and maps. The pattern 
\begin{babellisting}
{$p_1$, $\ldots$, $p_n$}
\end{babellisting}
matches a set that has $n$ elements $e_1$, $\ldots$, $e_n$ such that $p_i$ matches $e_i$ and $e_i < e_j$ for $i < j$. The pattern
\begin{babellisting}
{$p_1$, $\ldots$, $p_n$, $\delta$}
\end{babellisting}
matches a set that has $m \ge n$ elements such that its first $n$ elements match the patterns $p_i$, and the set consisting of the other $m-n$ elements matches the $\delta$-pattern. 

Similarly, the pattern
\begin{babellisting}
{$p_1$ -> $q_1$, $\ldots$, $p_n$ -> $q_n$}
\end{babellisting}
matches a map consisting of $n$ key/value pairs such that the key/value pairs match the pattern pairs in order.
The pattern \babelsrc{\{ -> \}} matches the empty map. Map patterns can have a $\delta$ pattern, too.

\section{Reals and Interval Arithmetic}\label{sec:reals}
Babel-17 is radical in its treatment of floating point arithmetic: there is \emph{only} interval arithmetic. What that means is that reals are represented in Babel-17 as closed real intervals, i.e. as pairs $[a; b]$ where $a$ and $b$ are floating point numbers and $a \leq b$. The usual floating point numbers $f$ can then be represented as $[f; f]$. For example, the notation $1.5$ is just short for $[1.5; 1.5]$.

The general form of this interval notation is
\begin{babellisting}
[$a$; $b$]
\end{babellisting}
Its value is formed by evaluating $a$ and $b$, resulting in \texttt{real}s $[a_1; a_2]$ and $[b_1; b_2]$, respectively, and then forming the interval $[\min(a_1, b_1), \max(a_2, b_2)]$. In case the evaluation does not yield reals, a \emph{DomainError}-exception is thrown.

The general rule of interval arithmetic in Babel-17 is as follows: If $f$ is a mapping from $\mathbb{R}^n$ to $\mathbb{R}$, then for its interval arithmetic implementation $F$ the following must hold for all $x_i \in \mathbb{R}$ and all \texttt{real}s $y_i = [u_i; v_i]$ such that $u_i \le x_i \le v_i$:
\[
f(x_1, \ldots, x_n) \in F(y_1, \ldots, y_n) 
\]
The built-in order on \texttt{real}s is defined such that
\[
[a_1; a_2] \leq [b_1; b_2] \quad\text{iff}\quad a_2 < b_1 \vee (a_1 = b_1 \wedge a_2 = b_2)
\]
Note that certain seemingly obvious inequalities like $[1.0; 2.0] \leq 2.0$ do \emph{not} hold! 

\section{Syntactic Sugar}
One of the goals of Babel-17 is that Babel-17 code is easy to read and understand. I have found that allowing arbitrary user-specific extensions to the syntax of code is definitely not helping to achieve this goal. Nevertheless, a bare minimum of syntactic sugar is necessary to support the most basic conventions known from arithmetic; for example, one should be able to write \texttt{3 + 5} to denote the addition of the integer 3 and the integer 5.

The programmer can use some of this syntactic sugar when defining her own objects. For example, \texttt{3 + 5} is just syntactic sugar for 
\begin{babellisting}
3.plus_ 5
\end{babellisting} 
Table~\ref{tab:syntacticsugar} lists all syntactic sugar of Babel-17 that is also available to the programmer. 
\begin{table}
\caption{Syntactic Sugar}
\begin{tabular}{c|c}
\textbf{Sugared} & \textbf{Desugared} \\\hline\hline
$a$ \verb!+! $b$ & \babelsrc{$a$.plus_ $\ b$} \\\hline
$a$ \verb!-! $b$ & \babelsrc{$a$.minus_ $\ b$} \\\hline
\verb!-! $a$ & \babelsrc{$a$.uminus_} \\\hline
$a$ \verb!*! $b$ & \babelsrc{$a$.times_  $\ b$} \\\hline
$a$ \verb!/! $b$ & \babelsrc{$a$.slash_  $\ b$} \\\hline
$a$ \verb!div! $b$ & \babelsrc{$a$.div_  $\ b$} \\\hline
$a$ \verb!mod! $b$ & \babelsrc{$a$.mod_  $\ b$} \\\hline
$a$ \verb+^+ $b$ & \babelsrc{$a$.pow_  $\ b$} \\\hline
$a$ \verb!++! $b$ & \babelsrc{$a$.plus__ $\ b$} \\\hline
$a$ \verb!--! $b$ & \babelsrc{$a$.minus__ $\ b$} \\\hline
$a$ \verb!**! $b$ & \babelsrc{$a$.times__  $\ b$}\\\hline
$a$ \verb!//! $b$ & \babelsrc{$a$.slash__  $\ b$}\\\hline
$a$ \verb!to! $b$ & \babelsrc{$a$.to_ $\ b$}\\\hline
$a$ \verb!downto! $b$ & \babelsrc{$a$.downto_  $\ b$}\\\hline 
$f$ $x$ &  \babelsrc{$f$.apply_  $\ x$}
\end{tabular}
\label{tab:syntacticsugar}
\end{table}
The availability of syntactic sugar for function application means that you can let your own objects behave as if they were functions. 

\section{Memoization}
Babel-17 supports \emph{memoization}. 
In those places where \babelsrc{def}-statements can be used, \babelsrc{memoize}-statements can be used, also. A \babelsrc{memoize}-statement must always refer to \babelsrc{def}-statements in the same scope. Babel-17 differentiates two kinds of memoization, \emph{strong} and \emph{weak}. A \babelsrc{memoize}-statement has the following syntax:
\begin{babellisting}
memoize $\metababel{ref}_1, \hdots, \metababel{ref}_n$
\end{babellisting}
The $\metababel{ref}_i$ are either of the form $\metababel{id}$ to indicate strong memoization, or of the form $(\metababel{id})$ to signal weak memoization. In both cases $\metababel{id}$ refers to the \babelsrc{def}-statement being memoized. As an example, here is the definition of the weakly memoized fibonacci-sequence:
\begin{babellisting}
memoize (fib)
def fib 0 = 0
def fib 1 = 1
def fib n = fib (n-1) + fib (n-2)
\end{babellisting}
The difference between weak and strong memoization is that strong memoization always remembers a value once it has been computed; weak memoization instead may choose not to remember computed results in certain situations, for example in order to free memory in low memory situations. Note that all arguments to the same memoized function must be totally ordered by the built-in partial order. 

\section{Linear Scope}
We are now ready to explore the full power of block expressions and statements in Babel-17. We approach the topic of this and the following sections, \emph{linear scope}, in an example-oriented way. A more in-depth treatment can be found in my paper \emph{Purely Functional Structured Programming}~\cite{pfsp}.

First, we extend the syntax of \babelsrc{val}-statements such that it is legal to leave out the keyword \babelsrc{val} in certain situations. Whenever the situation is such that it is legal to write
\begin{babellisting}
$a$ = $\textit{expr}$
\end{babellisting}
for a variable identifier $a$, we say that $a$ is in \emph{linear scope}. We call the above kind of statements \emph{assignments}. A necessary condition for $a$ to be in linear scope is for $a$ to be in scope because of a previous \babelsrc{val}-statement or a pattern match. 
The idea behind linear scope is that when the program control flow has a linear structure, then we can make assignments to variables in linear scope without leaving the realm of purely functional programming.

The following two expressions are equivalent:
\begin{center}
\begin{tabular}{ccc}
\begin{babellisting}
begin
  val x = $a$ 
  val y = $b$
  val x = $c$
  $d$
end
\end{babellisting}
& \hspace{2cm} &
\begin{babellisting}
begin
  val x = $a$ 
  val y = $b$
  x = $c$
  $d$
end
\end{babellisting}
\end{tabular}
\end{center}
So far, \babelsrc{val}-statements and assignments have indistinguishable semantics. The differences start to show when we look at nested block expressions:
\begin{center}
\begin{tabular}{ccccc}
\begin{babellisting}
begin
  val x = 1 
  val y = 2
  begin
    val x = 3
    val y = 4 * x
  end
  (x, y)
end
\end{babellisting}
& \hspace{1cm} &
\begin{babellisting}
begin
  val x = 1 
  val y = 2
  begin
    val x = 3
    y = 4 * x
  end
  (x, y)
end
\end{babellisting}
& \hspace{1cm} &
\begin{babellisting}
begin
  val x = 1 
  val y = 2
  begin
    val x = 3
    val y = 0
    y = 4 * x
  end
  (x, y)
end
\end{babellisting}\\\hline
evaluates to $(1, 2)$ & &
evaluates to $(1, 12)$ & &
evaluates to $(1, 2)$ \\[0.5cm]
\end{tabular}
\end{center}
The above examples show that the effect of dropping the \babelsrc{val} in a \babelsrc{val} statement is that the binding of an identifier becomes visible at that level within the linear scope where it has last been introduced via a \babelsrc{val} statement or via a pattern match.

Linear scope spreads along the statements and nested block expressions of a block expression. It usually does not spread into expressions. An exception are certain expressions that can also be viewed as statements. We call these expressions \emph{control expressions}. For example, in
\begin{babellisting}
begin
  val x = 1 
  x = 2
  val y = $a$ * $b$
 (x, y)
end
\end{babellisting}
the linear scope of $x$ does not extend into the expressions $a$ and $b$, because $a * b$ is not a control expression. Therefore the above expression will always evaluate (assuming there is no exception) to a pair which has $2$ as its first element. But for example in
\begin{babellisting}
begin
  val x = 1 
  x = 2
  val y = 
     begin
        $s_1$
        $\vdots$
        $s_n$
     end
  (x, y)
end
\end{babellisting}
the linear scope of $x$ extends into $s_1$ and spreads then along the following statements. Therefore the first element of the pair that is the result of evaluating above expression depends on what happens in the $s_i$.
Here are three example code snippets that further illustrate linear scope:
\begin{center}
\begin{tabular}{ccccc}
\begin{babellisting}
begin
  val x = 1
  val y = 
    begin 
      x = 2 
      x+x 
    end
  (x, y)
end
\end{babellisting}
& \hspace{1cm} &
\begin{babellisting}
begin
  val x = 1
  val y = 3 *
    begin 
      x = 2 
      x+x 
    end
  (x, y)
end
\end{babellisting}
& \hspace{1cm} &
\begin{babellisting}
begin
  val x = 1
  val y = 3 *
    begin 
      val x = 2 
      x+x 
    end
  (x, y)
end
\end{babellisting}\\\hline
evaluates to $(2, 4)$ & &
illegal & &
evaluates to $(1, 12)$ \\[0.5cm]
\end{tabular}
\end{center}

These are the control expressions that exist in Babel-17:
\begin{itemize}
\item \babelsrc{begin ... end}
\item \babelsrc{if ... end}
\item \babelsrc{match ... end}
\item \babelsrc{try ... catch ... end}
\item \babelsrc{for ... do ... end}
\item \babelsrc{while ... end}
\end{itemize}
The last two control expressions are loops and explained in the next section. The first four have already been treated without delving too much into their statement character. We have already seen how  \babelsrc{begin ... end} is responsible for nesting block expressions, when it is used as a statement.  Just as  the \babelsrc{begin ... end} expression may be used as a statement, 
you can also use all other control expressions as statements, for example:
\begin{babellisting}
begin
  val x = random 2
  if x == 0 then
    x = 100
  else
    x = 200
  end
  x + x
end
\end{babellisting}
This expression will evaluate either to $200$ or to $400$.

For \babelsrc{if}-statements the \babelsrc{else}-branch is optional, but \babelsrc{match}-statements throw a \babelsrc{NoMatch} exception if none of the patterns matches. 

\section{Record Updates}
Linear scope makes it possible to have purely functional record updates in Babel-17. Let us assume you have defined $u$ via
\begin{babellisting}
val u = {x = 10, y = 20, z = -4}
\end{babellisting}
and now you want to bind $u$ to another record that differs only in the $x$-component. You could proceed as follows:
\begin{babellisting}
u = {x = 9, y = u.y, z = u.z}
\end{babellisting}
but this clearly does not scale with the number of components of $u$. A more scalable alternative would be to write
\begin{babellisting}
u = 
  object + [u]
    def x = 9
  end      
\end{babellisting}
Babel-17 allows to write down the above statement in a more concise form:
\begin{babellisting}
u.x = 9   
\end{babellisting}
In general, for a value $u$ of type \texttt{obj},
\begin{babellisting}
u.$m$ = $t$   
\end{babellisting}
is shorthand notation for 
\begin{babellisting}
u = 
  begin
    val evaluated_t = $t$
    object + [u]
      def $m$ = evaluated_t
    end
  end      
\end{babellisting}

Actually, there is an exception to the above rule: if $u$ has a message \texttt{$m$\_putback\_}, then 
\begin{babellisting}
u.$m$ = $t$   
\end{babellisting}
is actually shorthand for 
\begin{babellisting}
u = u.$m$_putback_ $t$
\end{babellisting}

The reason for this exception is that it allows us to generalize the concept of a purely functional record update to the concept of a \emph{lens}.

\section{Lenses}
Lenses~\cite{lenses} allow us to generalize the thing we did with updating records in a purely functional way. Lenses and linear scope work together well; this is very similar to how lenses and the state monad work together well~\cite{lensesfi}.

A \emph{lens} is basically a pair of functions $(g, p)$, where $g$ is called the \emph{get} function and $p$ is called the \emph{putback} function of the lens. In the special case where a lens represents a field $m$ of a record $u$, this pair would be defined via
\[
	g\ u= u.m, \quad p\ u\ t = \textsl{a copy of $u$ where the field $m$ has been set to $t$}
\]
Generalizing from this special case, a lens should obey the following laws:
\[
	g\ (p\ u\ t) = t,\quad p\ u\ (g\ u) = u
\]
In Babel-17, these laws are just recommendations for how to construct a lens as there are no mechanisms in place to ensure that a lens actually obeys them.

You can define a lens in Babel-17 directly via providing \emph{get} and \emph{putback} explictly:
\begin{babellisting}
def g u = u.m
def p u = t => begin u.m = t; u end
val l = lens (g, p)
\end{babellisting}
The above code constructs a new lens $l$. Lenses actually form their own type, so we have \babelsrc{typeof l == (: lens_)}.

What can you do with a lens once you got one? You can apply $l$ to a value $u$ via
\begin{babellisting}
l u
\end{babellisting}
This applies the \emph{get} function of $l$ to $u$, so for our specific $l$ we have that \babelsrc{l u} is equivalent to \babelsrc{u.m}.
In order to emphasize the aspect that $l$ can be regarded as a field accessor for $u$, instead of \babelsrc{l u} you can also write
\begin{babellisting}
u.(l)
\end{babellisting}
This notation has the advantage that it can also be used on the left hand side of an assignment:
\begin{babellisting}
val u = {m = 10, n = 12}
u.(l) = 23
u == {m = 23, n = 12}
\end{babellisting}
The above evaluates to \babelsrc{true}, because the line \babelsrc{u.(l) = 23} is short for \babelsrc{u = l.putback u 23}. 

Lenses come automatically equipped with a \emph{modify} operation, \babelsrc{l.modify u f} is short for \babelsrc{l.putback u (f (l u))}. There is special notation that supports this operation:
\begin{babellisting}
val u = {m = 10, n = 12}
u.(l) += 2
u == {m = 12, n = 12}
\end{babellisting}
This again evaluates to \babelsrc{true}, because the line \babelsrc{u.(l) += 2} is short for 
\begin{babellisting}
u = l.modify u (m => m +  2)
\end{babellisting}
The \emph{identity lens} 
\begin{babellisting}
val id = lens (x => x, x => y => y)
\end{babellisting}
blurs the distinction between normal assignment and lens assignment, because the meaning of
\begin{babellisting}
val x = 10
x = 12
\end{babellisting}
is exactly the same as the meaning of
\begin{babellisting}
val x = 10
x.(id) = 12
\end{babellisting}
In particular, our new assignment operators like \babelsrc{+=} and \babelsrc{*=} also work in the context of normal assignment:
\begin{babellisting}
val x = 10
x += 2
x == 12
\end{babellisting}
This evaluates to \babelsrc{true}. Tables~\ref{table:modifyingoperators} list all modifying assignment operators.
\begin{table}
\caption{Modifying Assignment Operators}
\begin{tabular}{r|l}
\textbf{Syntax} & \textbf{Semantics} \\\hline
x += y & x = x + y \\
x =+ y & x = y + x \\
x ++=y & x = x ++ y\\
x =++ y & x = y ++ x\\
x $-$= y & x = x $-$ y \\
x =$-$ y & x = y $-$ x \\
x $-$$-$= y & x = x $-$$-$ y\\
x =$-$$-$ y & x = y $-$$-$ x\\
x xor= y & x = x xor y \\
x =xor y & x = y xor x \\
\end{tabular}
\begin{tabular}{r|l}
\textbf{Syntax} & \textbf{Semantics} \\\hline
x *= y & x = x * y \\
x =* y & x = y * x \\
x **=y & x = x ** y\\
x =** y & x = y ** x\\
x /= y & x = x / y \\
x =/ y & x = y / x \\
x //= y & x = x // y\\
x =// y & x = y // x\\
x and= y & x = x and y \\
x =and y & x = y and x \\
\end{tabular}
\begin{tabular}{r|l}
\textbf{Syntax} & \textbf{Semantics} \\\hline
x \verb+^+= y & x = x \verb+^+ y \\
x =\verb+^+ y & x = y \verb+^+ x \\
x div= y & x = x div y \\
x =div y & x = y div x \\
x mod= y & x = x mod y \\
x =mod y & x = y mod x \\
x min= y & x = min (x, y) \\
x max= y & x = max (x, y) \\
x or= y & x = x or y \\
x =or y & x = y or x \\
\end{tabular}
\label{table:modifyingoperators}
\end{table}

Besides via directly giving the pair of functions that form a lens, a lens can also be defined by its \emph{access path}. The access path of a lens is basically just how you would write down the \emph{get} function of the lens if it could be defined using message passing (possibly with arguments) and lens application only. For example, instead of defining a lens via
\begin{babellisting}
def g u = u.m
def p u = t => begin u.m = t; u end
val l = lens (g, p)
\end{babellisting}
you could just write
\begin{babellisting}
val l = lens u => u.m
\end{babellisting}
The identity lens becomes even simpler in this form:
\begin{babellisting}
val id = lens u => u
\end{babellisting}
Another valid legal lens definition is:
\begin{babellisting}
val h = lens x => ((x.mymap.lookup 10) + 40).(l)
\end{babellisting}
You could define the above lens directly via a pair of functions as follows:
\begin{babellisting}
def g x = ((x.mymap.lookup 10) + 40).(l)
def p x = t => 
  begin
    val u1 = x.mymap
    val u2 = u1.lookup 10
    val u3 = u2 + 40
    val new_u3 = l.putback u3 t
    val new_u2 = u2.plus__putback_ 40 new_u3
    val new_u1 = u1.lookup_putback_ 10 new_u2
    val new_x = x
    new_x.mymap = new_u1
    new_x
  end

val h = lens (g, p)
\end{babellisting}

The cool thing about lenses is that they are \emph{composable}. Given two lenses $a$ and $b$, we can define a new lens $c$ via
\begin{babellisting}
val c = lens x => x.(a).(b)
\end{babellisting}
Multiplication of lenses is defined to be just this composition, so we could also just write
\begin{babellisting}
val c = a * b
\end{babellisting}

Here is an example that illustrates composition:
\begin{babellisting}
val a = lens x => x.a
val b = lens x => x.b

val x = {a = {a = 1, b = 2}, b = {a = 3, b = 4}}

x.(a*b) = 10
x.(b*a) = 20

x == {a = {a = 1, b = 10}, b = {a = 20, b = 4}}
\end{babellisting}
This evaluates to \babelsrc{true}. Note that if the left hand side of an assignment is an access path, then the corresponding lens is automatically constructed. Therefore, in the above context the following five lines have identical meaning:
\begin{babellisting}
x.(a*b) = 10
x.(a).(b) = 10
x.a.b = 10
x.(a).b = 10
x.a.(b) = 10
\end{babellisting}

\section{With}\label{sec:collector}
By default, all yielded values of a block expression are collected into a vector which collapses in the case of a single element. The programmer might want to deviate from this default and collect the yielded values differently, for example to get rid of the collapsing behavior. The \babelsrc{with} expression allows her to do just that. Its syntax is:
\begin{babellisting}
with $c$ do 
  $b$
end
\end{babellisting}
where $c$ is a \emph{collector} and $b$ is a block expression. A collector $c$ is any object that 
\begin{itemize}
\item responds to the message \verb+collector_close_+,
\item and returns via $c$.\verb+collector_add_+ $x$ another collector.
\end{itemize}
It is recommended that collectors also support the message \texttt{empty} to represent the empty collector that has not collected anything yet.
 
Lists, vectors, sets, maps and strings are built-in collectors which the programmer can use out-of-the-box; apart from that she can implement her own collectors, of course.

Here is an example where we use a set as a collector:
\begin{babellisting}
with {4} do
  yield 1
  yield 2
  yield 1
  10
end
\end{babellisting}
Above expression evaluates to \babelsrc{\{1, 2, 4, 10\}}.

\section{Loops}\label{sec:loops}
This is the syntax for the \babelsrc{while}-loop:
\begin{babellisting}
while $c$ do
  $b$
end
\end{babellisting}
Here $c$ must evaluate to a boolean and $b$ is a block expression. For example, here is how you could code the euclidean algorithm for calculating the greatest common divisor:
\begin{babellisting}
def gcd (a,b) = begin
  while b <> 0 do
    (a, b) = (b, a mod b)
  end
  a
end
\end{babellisting}

There is also the \babelsrc{for}-loop. It has the following syntax:
\begin{babellisting}
for $p$ in $C$ do
  $b$
end
\end{babellisting}
In the above $p$ is a pattern, $C$ is a \emph{collection}, and $b$ is a block expression.
The idea is that above expression iterates through those elements of the \emph{collection} $C$ 
which match $p$; for each successfully matched element, $b$ is executed. 

An object $C$ is a collection if it handles the message \texttt{iterate\_}
\begin{itemize}
\item by returning \texttt{()} if it represents the empty collection,
\item or otherwise by returning \texttt{($e$, $C'$)} such that $C'$ is also a collection.
\end{itemize}

Here is an example of a simple \babelsrc{for}-loop expression:
\begin{babellisting}
begin
  val s = [10, (5, 8), 7, (3,5)]
  with {->} : for (a,b) in s do
    yield (b,a)
  end
end
\end{babellisting}
evaluates to \babelsrc{\{8 -> 5, 5 -> 3\}}.

Using \babelsrc{for}-loops in combination with linear scope it is possible to formulate all of those \emph{fold}-related functionals known from functional programming in a way which is easier to parse (and remember) for most people. Let us for example look at a function that takes a list $m$ of integers $[a_0, \ldots, a_n]$ and an integer $x$ as arguments and returns the list
\begin{displaymath}
	[q_0, \ldots, q_n] \quad \text{where} \quad q_k = \sum_{i=0}^k a_i\, x^i
\end{displaymath}
The implementation in Babel-17 via a loop is straightforward, efficient and even elegant:
\begin{babellisting}
m => x => 
  with [] do 
    val y = 0
    val p = 1
    for a in m do 
      y = y + a*p
      p = p * x
      yield y
    end
 end
\end{babellisting}

The built-in collections that can be used with for-loops are the usual suspects: lists, vectors, sets, maps and strings. Of course you can define your own custom collections. There is even a pattern you can use for matching against an arbitrary collection:
\begin{babellisting}
(for $p_1$, $\ldots$, $p_n$ end)
\end{babellisting}
and the corresponding $\delta$ pattern
\begin{babellisting}
(for $p_1$, $\ldots$, $p_n$, $\delta$ end)
\end{babellisting}
match collections with exactly $n$ or at least $n$ elements, respectively. 

\section{Pragmas}
Pragmas are statements that are not really part of the program, but inserted in it for pragmatical reasons. They are useful for testing, debugging and profiling a program. There are currently four different kinds of  pragmas available:

\texttt{\#log $e$} evaluates the expression $e$ and logs it. 

\texttt{\#print $e$} evaluates the expression $e$ such that it has no internal lazy or concurrent computations any more, and then logs it.

\texttt{\#profile $e$} evaluates the expression $e$, gathers profiling information while doing so, and logs both.

\texttt{\#assert $e$} evaluates the expression $e$ and signals an error if $e$ does not evaluate to \babelsrc{true}.

\texttt{\#catch $p$ try $e$} evaluates the expression $e$ and signals an error if $e$ does not evaluate to an exception with a parameter that matches $p$.

The Babel-17 interpreter / compiler is free to ignore pragmas, typically if instructed so by the user. 

\section{Modules and Types}\label{modules}
Modules give Babel-17 code a static structure and improve on object expressions by providing more sophisticated encapsulation mechanisms via types. Modules are introduced with the syntax
\begin{babellisting}
module $p_1.p_2.\ldots.p_n$
  $s_1$
  $\vdots$
  $s_n$  
end
\end{babellisting}
where $p = p_1.p_2.\ldots.p_n$ is the \emph{module path}. 

The statements $s_i$ that may appear in modules are basically those that may also appear in object expressions. Additionally, modules can contain \emph{type definitions}; this is the topic of the next section. 

Identifiers introduced via definitions become the messages that this module responds to except when hidden via \babelsrc{private}. 

Modules can be nested. Basically,
\begin{babellisting}
module $p_1.p_2.\ldots.p_n$
  $S_1$
  module $q_1.q_2.\ldots.q_m$
  $\vdots$
  end  
  $S_2$  
end
\end{babellisting}
means the same as
\begin{babellisting}
module $p_1.p_2.\ldots.p_n$
  $S_1$
  $S_2$  
end
module $p_1.p_2.\ldots.p_n.q_1.q_2.\ldots.q_m$
$\vdots$
end  
\end{babellisting}
A module cannot access its surrounding scope, with the exception of those identifiers that have been introduced via \babelsrc{import}. So the following is not allowed, because inside module $b$ the identifier $u$ cannot be accessed:
\begin{babellisting}
module $a$
  val u = 2
  module b
    def message = u
  end  
end
\end{babellisting}
But the following \emph{is} allowed:
\begin{babellisting}
module $a$
  import someCoolModule.importantValue => u
  module b
    def message = u
  end  
end
\end{babellisting}

\subsection{Loading Modules}
Before a module can respond to messages, it must be loaded. 
A module is in one of the following three states: DOWN, LOADING or UP. There are three things that one might  suspect to change the state of the module: 
\begin{description}
\item[A] the module receives a message $m$ that corresponds to a definition inside of it,
\item[B] the module receives a message $m$ that corresponds to a submodule $p.m$,
\item[C] the module receives an invalid message $m$.
\end{description}
Of these three things, neither B nor C change the state of the module, so we are looking now only at case A.

In the following we distinguish modules which need initialization from modules which don't. A module does not need initialization iff it consists only of the following types of statements: \babelsrc{def}, \babelsrc{typedef}, \babelsrc{import}, \babelsrc{private}, \babelsrc{memoize}. 

If module $p$ needs initialization and is in state DOWN, then sending it a message causes it to go to state LOADING; after $p$ has been loaded, that is all statements of the module have been executed, $p$ goes from state LOADING to state UP. If the module needs no initialization, the state goes directly from DOWN to UP.

If $p$ is in state LOADING, then sending it a message will make the sender wait until $p$ is in state UP. Note that this can lead to a deadlock situation. In case the Babel-17 runtime can detect this deadlock, a \emph{DeadLock}-exception is thrown. 

If $p$ is already in state UP, then A does not change the state of the module. 

Note that the state of a module is not directly accessible and only sometimes indirectly observable via the \emph{DeadLock}-exception.

Here is an example that leads to a deadlock:
\begin{babellisting}
module deadlock
  def x = 10
  val a = deadlock.x + 1
  def y = a * a
end
deadlock.y
\end{babellisting}
Note that the following example does \emph{not} lead to a deadlock:
\begin{babellisting}
module noDeadlock
  def x  = 10
  val a = x + 1
  def y = a * a
end
noDeadlock.y
\end{babellisting}
The most straightforward way to avoid deadlock problems is to write modules which need no initialization. 

\subsection{Type Definitions}\label{sec:typedefs}
A module may contain type definition statements. A new type $t$ is introduced via
\begin{babellisting}
typedef $t$ $pat$ = $expr$
\end{babellisting}
where $t$ is an identifier, $pat$ a pattern and $expr$ an expression. Like a \babelsrc{def} statement, the above introduces a new function $t$. Unlike a \babelsrc{def} statement, it also introduces a new type $t$. The new function $t$ has the property that all non-exceptional values $v$ it returns will have type $t$ and will consist of two components: an \emph{inner value}, and an \emph{outer value}. The inner value $i$ is just the argument that was passed to $t$. The outer value $o$ is the result of evaluating and forcing $expr$ in case this does not lead to an exception. If $pat$ does not match, then a \emph{DomainError}-exception is thrown. 

The point of all this is that the value $v$ behaves just like its outer value $o$, i.e. $v$ responds to the same messages as $o$, and does so in the same way. Nevertheless, the type of $v$ is $t$, whatever the type of $o$ might be. This means that you can implement the behavior of $v$ by choosing an appropriate $o$ which accesses the hidden encapsulated state $i$. This hidden state can also be accessed in the module in which the type has been defined via the \emph{inner-value pattern} ($t$\ $p$) which matches $v$ if it has type $t$ and its inner value $i$ matches $p$.

Let's look at a first type definition example:
\begin{babellisting}
module cards
  typedef rank i = 
    match i 
      case 14 => Ace
      case 13 => King
      case 12 => Queen
      case 11 => Jack
      case x if 2 <= x <= 10 => Number x
    end
end
\end{babellisting}
In the above we define the type \emph{cards.rank}. We can create a value of this type like this:
\begin{babellisting}
val k = cards.rank 13
\end{babellisting}
The expression \babelsrc{(:cards.rank)}\ evaluates to the type \emph{cards.rank}, therefore 
\begin{babellisting}
(typeof k) == (:cards.rank)
\end{babellisting}
evaluates to \babelsrc{true}. 

Note that there is no other way to create a value of type \emph{cards.rank} than via the \emph{rank} function in the module \emph{cards}.  If you want to access the inner value of a value of type \emph{cards.rank}, you can do so \emph{within} the module \emph{cards} via the inner-value pattern:
\begin{babellisting}
module cards
  typedef rank ...
  ...
  def rank2num(rank n) = n
  ...
end
\end{babellisting}
Now you can access the inner value of a rank even outside of the \emph{cards} module by calling the \emph{rank2num} function we just defined. The following evaluates to \babelsrc{true}:
\begin{babellisting}
cards.rank2num (cards.rank 14) == 14
\end{babellisting}
You can use the fact that a rank value behaves like its outer value to do pattern matching on rank values:
\begin{babellisting}
def rank2str(r : rank) = 
  match r
    case (Ace !) => "ace"
    case (King !) => "king"
    case (Queen !) => "queen"
    case (Number ! n) => "number"+n
  end
\end{babellisting}
It is possible to distribute the definition of a type $t$ over several statements:
\begin{babellisting}
typedef $t$ $pat_1$ = $expr_1$
  $\vdots$
typedef $t$ $pat_n$ = $expr_n$
\end{babellisting}
So an alternative way of defining the rank type would have been:
\begin{babellisting}
module cards
  typedef rank 14 = Ace
  typedef rank 13 = King
  typedef rank 12 = Queen
  typedef rank 11 = Jack
  typedef rank (n if 2 <= n <= 10) = Number n
end
\end{babellisting}
As a shortcut notation, several typedef clauses belonging to the same type $t$ can be combined into a single one by separating the different cases by commas:
\begin{babellisting}
module cards
  typedef rank 14 = Ace, 13 = King, 12 = Queen, 11 = Jack, 
                (n if 2 <=n <= 10) = Number n
end
\end{babellisting}
Another shortcut notation allows you to write instead of
\begin{babellisting}
typedef $t$ (x as $p$) = x
\end{babellisting}
just 
\begin{babellisting}
typedef $t$ $p$
\end{babellisting}

Finally, it is possible to identify a type with the module it is defined in by using the same name for the type and the module. For example:
\begin{babellisting}
module util.orderedSet
  typedef orderedSet ...
...
end
\end{babellisting}
This defines the module \emph{util.orderedSet} and the type \emph{util.orderedSet.orderedSet}. Because of our convention of identifying modules and types that have the same name, you can refer to the type \emph{util.orderedSet.orderedSet} also just by \emph{util.orderedSet}. In particular, the following evaluates to \babelsrc{true}:
\begin{babellisting}
(:util.orderedSet.orderedSet) == (:util.orderedSet)
\end{babellisting}

This is all there is to defining your own types in Babel-17. In the following we list a few common type definition patterns: \emph{simple enumerations}, \emph{algebraic datatypes}, and \emph{abstract datatypes}.

\subsection{Simple Enumerations}
It is easy to define a type that is just an enumeration:
\begin{babellisting}
module cards
  typedef suit Spades, Clubs, Diamonds, Hearts
end
\end{babellisting}
We may choose to represent our rank type rather as an enumeration, too:
\begin{babellisting}
module cards
  typedef rank Ace, King, Queen, Jack, Number (n if 2 <= n <= 10)
end
\end{babellisting}

\subsection{Algebraic Datatypes}
For enumeration types, the inner value is always equal to the outer value. This property is also shared by the more complex algebraic datatypes which are typically recursively defined: 
\begin{babellisting}
module cards
  typedef bintree Leaf _, Branch ( _ : bintree, _ : bintree)
end
\end{babellisting}

\subsection{Abstract Datatypes}
The type definition facility of Babel-17 is powerful enough to allow you to define your own abstract datatypes. Currently there is no type in Babel-17 for representing sets that are ordered by a given (i.e., not necessarily the built-in) order. So let's define our own. To keep it simple, we use lists to internally represent sets:
\begin{babellisting}
module util.orderedSet

  private orderedSet, ins
  
  typedef orderedSet (leq, list) = nil
  
  def empty (leq : fun) = orderedSet (leq, [])
  
  def insert (orderedSet (leq, list), y) = (orderedSet (leq, ins (leq, list, y))
  
  def toList (orderedSet (leq, list)) = list
  
  def ins (leq, [], y) = [y]
  def ins (leq, x::xs, y) =
    if leq (y, x) then
        if leq (x, y) then x::xs else y::x::xs end
    else
        x::(ins (leq, xs, y))
    end  
end
\end{babellisting}
From outside the module, the only way to create values of type \emph{orderedSet}  is by calling \emph{empty} and \emph{insert}. The only way to inspect the elements of an \emph{orderedSet} is via \emph{toList}.

\subsection{Type Conversions}
It often makes sense to convert a value of one type into a value of another type. Probably the most prominent example of this is converting an \emph{Integer} to a \emph{Real} and rounding a \emph{Real} to an \emph{Integer}.

Because type conversions are so ubiquitous, there is special notation for it in Babel-17. To convert a value $v$ canonically to a value of type $t$, the notation
\begin{babellisting}
$v$ :> $t$
\end{babellisting}
is used. If $t$ is a value of type \texttt{type}, you can write
\begin{babellisting}
$v$ :> ($t$)
\end{babellisting}
to express the conversion operation.

Your own object can implement conversion to a type $t$ via 
\begin{babellisting}
object 
  ...
  def this :> $t$ = ...
  ....
end
\end{babellisting}
In case your object cannot convert to $t$, it should throw a \emph{DomainError}-exception.

You can annotate your \emph{def}-statements with a \emph{return type} $t$, for example:
\begin{babellisting}
def myfun $pat$ : $t$ = $expr$
\end{babellisting}
This is just another notation for
\begin{babellisting}
def myfun $pat$ = ($expr$ :> $t$)
\end{babellisting}

\subsection{Automatic Type Conversions}\label{sec:automatictypeconversion}
Type conversions are actually differentiated into two separate classes: those which are automatically applied, and those which are only applied in conjunction with the \texttt{:>} operator.

Your own object can implement automatic type conversion to a type $t$ via 
\begin{babellisting}
object 
  ...
  def this : $t$ = ...
  ....
end
\end{babellisting}
Note that an automatic type conversion takes precedence over a non-automatic one when used in conjunction with the \texttt{:>} operator. In particular,
\begin{babellisting}
object 
  def this : int = 1
  def this :> int = 2
end :> int
\end{babellisting}
will evaluate to $1$.

\subsection{Import}
A module can be accessed from anywhere via its module path. For example will 
\begin{babellisting}
module hello.world
  def x = 2
end
hello.world.x
\end{babellisting}
result in the value $2$.

To avoid typing long module paths you can \emph{import} them:
\begin{babellisting}
import hello.world.x
(x, x, x)
\end{babellisting}
will evaluate to $(2, 2, 2)$. 

You can also import types this way:
\begin{babellisting}
import util.orderedSet.orderedSet
def e : orderedSet = util.orderedSet.empty
\end{babellisting}
Alternatively, to make the above statement even shorter, import \emph{both} the type and its enclosing module:
\begin{babellisting}
import util.orderedSet
def e : orderedSet = orderedSet.empty
\end{babellisting}
It is not possible to just import all members of a module; Babel-17 is dynamically typed, and the danger of accidental mayhem just would be too big.

You can rename imports, for example:
\begin{babellisting}
import util.orderedSet => set
def e : set = set.empty
\end{babellisting}
You can combine several imports into a single one:
\begin{babellisting}
import util.orderedSet.{empty, insert => orderedInsert} 
def e : set = set.empty
\end{babellisting}
All imports must be grouped together at the beginning of a block, and later \babelsrc{import}-statements in this group take earlier \babelsrc{import}-statements in this group (and outside this group, of course) into account.

There is the \babelsrc{root}-keyword that denotes the module root.  Instead of \babelsrc{import util.orderedSet} you could just as well say
\begin{babellisting}
import root.util.orderedSet
\end{babellisting}
You can also use it in expressions, like in
\begin{babellisting}
val e = root.util.orderedSet.empty
\end{babellisting}
The \babelsrc{root}-keyword comes in handy in situations when you want to explicitly make sure that there is no name aliasing going on:
\begin{babellisting}
import com.util
import util.orderedSet
\end{babellisting}
here the second import actually imports \babelsrc{com.util.orderedSet}. If you want to import \babelsrc{util.orderedSet} instead, write
\begin{babellisting}
import com.util
import root.util.orderedSet
\end{babellisting}

\subsection{Abstract Datatypes, Continued}
We have seen previously how to define our own abstract datatype \emph{orderedSet}. This is already fine, but we would like orderedSet to integrate more tightly with how things work in Babel-17. In particular,
\begin{enumerate}
\item ordered sets should be comparable in a way that makes sense; right now, the expression
\begin{babellisting}
insert (empty leq, 1) == insert (empty leq, 2)
\end{babellisting}
evaluates to \babelsrc{true} because both values have outer value \babelsrc{nil},
\item instead of \babelsrc{insert (r, x)} we just want to write \babelsrc{r + x},
\item instead of \babelsrc{toList r} we want to write \babelsrc{r :> list},
\item we want to be able to traverse the set via
\begin{babellisting}
for x in r do ... end
\end{babellisting}
\item and we want to be able to write 
\begin{babellisting}
with empty leq do
  yield a
  yield b
  yield c
end
\end{babellisting}
for the set consisting of the elements $a$, $b$ and $c$.
\end{enumerate}
We can achieve all that by replacing the line 
\begin{babellisting}
typedef orderedSet (leq, list) = nil
\end{babellisting}
with the following code:
\begin{babellisting}
typedef orderedSet (leq, list) = 
  object

    ## (1)
    def compare_ (orderedSet (leq2, list2)) = list $\sim$ list2

    ## (2)
    def plus_ x = insert (this, x)

    ## (3)
    def this : list = list

    ## (4)
    def iterate_ =
        match list
            case [] => ()
            case (x::xs) => (x, orderedSet (leq, xs))
        end

    ## (5)
    def collector_close_ = this
    def collector_add_ x = this + x
    def empty = orderedSet (leq, [])

  end
\end{babellisting}

\section{Unit Tests}
Unit testing has become a corner stone of most popular software development methodologies. Therefore Babel-17 provides language support for unit testing in form of the \babelsrc{unittest} keyword. This keyword can be used in two forms. 

First, it can be used as part of the name of a module, for example like in
\begin{babellisting}
module util.orderedSet.unittest

import util.orderedSet._

def leq (a, b) = a <= b
def geq (a, b) = a >= b

#assert insert (insert (empty leq, 3), 1) :> list == [1, 3]
#assert insert (insert (empty geq, 3), 1) :> list == [3, 1]

end
\end{babellisting}
Second, it can be used to separate a module definition into two parts:
\begin{babellisting}
module util.orderedSet

typedef orderedSet ...

...

unittest

def leq (a, b) = a <= b
def geq (a, b) = a >= b

#assert insert (insert (empty leq, 3), 1) :> list == [1, 3]
#assert insert (insert (empty geq, 3), 1) :> list == [3, 1]

end
\end{babellisting}
The meaning of the above is
\begin{babellisting}
module util.orderedSet

typedef orderedSet ...

...

def unittest = 
  begin

    def leq (a, b) = a <= b 
    def geq (a, b) = a >= b

    #assert insert (insert (empty leq, 3), 1) :> list == [1, 3]
    #assert insert (insert (empty geq, 3), 1) :> list == [3, 1]

  end

end
\end{babellisting}
except for the fact that this is not legal Babel-17.

The idea is that Babel-17 can run in two modes. In \emph{production mode}, all unit tests are ignored. In \emph{unit testing mode}, you can run your unit tests. Usually in production mode, you would switch off assertions, and in unit testing mode you would turn them on, but that is up to you.

An important detail is that production code is separated from unit testing code, i.e. while unit testing code can import other unit testing code and production code, production code can only import other production code. For example,
\begin{babellisting}
module mystuff

import util.orderSet.unittest => test

test ()

end
\end{babellisting}
is not legal Babel-17, because the production code module \emph{mystuff} cannot import the unit testing function \emph{util.orderSet.unittest}. On the other hand, the following is valid Babel-17:
\begin{babellisting}
module mystuff.unittest

import util.orderSet.unittest => test

test ()

end
\end{babellisting}

\section{Native Interface}
Babel-17 exposes the platform it is running on through its \emph{native interface}. To query which platform you are running on, use the expression
\begin{babellisting}
native Platform
\end{babellisting}
If there is no underlying platform you can access, the value \babelsrc{nil} is returned.
\subsection{The Java Platform}
In case of
\begin{babellisting}
native Platform == Java
\end{babellisting}
there is a limited form of Java interoperability available. With
\begin{babellisting}
val v = native New ($classname$, $x_1$, $\ldots$, $x_n$)
\end{babellisting}
you can create a value $v$ of type \verb+native_+ that is a wrapper for a Java object of class $classname$; the $x_i$ are the arguments that are passed to the constructor. Table~\ref{table:javaconversion} describes how values are converted between Babel-17 and Java.
\begin{table}
\caption{Conversions between Babel-17 and Java Values}
\begin{tabular}{r|l}
\textbf{Babel-17} & \textbf{Java} \\\hline
\babelsrc{nil} & \texttt{null} \\
\babelsrc{native_} & \texttt{Object} \\
\texttt{int} & \patterndescr {\texttt{byte Byte short Short char Char int Integer long Long BigInteger}} \\
\texttt{real} & \texttt{float Float double Double} \\
\texttt{bool} & \texttt{boolean Boolean} \\
\texttt{string} & \texttt{String char Char} \\ 
\texttt{vect} & Java array \\
\end{tabular}
\label{table:javaconversion}
\end{table}

You can access the fields of $v$ and call its methods by sending messages to $v$. This holds also true for static fields and messages. So for example, you can do the following:
\begin{babellisting}
val ar = native New ("java.util.ArrayList", 5)
val _ =
  begin
    ar.add 1
    ar.add 10
    ar.add 5
    ar.add ((18, 13, 15),)
  end
ar.toArray ()
\end{babellisting}
This evaluates to \texttt{(1, 10, 5, (18, 13, 15))}.

Ambiguities are resolved in \emph{some} way. The only rule you can rely on is that access to methods shadows access to fields. Note that ambiguities not only arise because of typing ambiguities, but also because messages in Babel-17 are case-insensitive, but in Java method and field names are case-sensitive.

You can instantiate a native value that corresponds to a Java class $classname$ via
\begin{babellisting}
val c = native Class $classname$
\end{babellisting}
The value $c$ will in addition to the normal class fields and methods also understand messages that correspond to the static fields and methods of the class. For example:
\begin{babellisting}
val c = native Class "java.lang.Integer"
c.parseInt "120"
\end{babellisting}
will evaluate to 120.

\section{Standard Library}
The standard library is an important part of the language definition of Babel-17 and consists of the built-in types of Babel-17. In future versions of Babel-17, also the module \emph{lang} and all its submodules are part of the standard library.

The messages implemented by the built-in types of the standard library can be grouped as follows:
\begin{itemize}
\item collector related messages (Section~\ref{sec:collector}),
\item collection related messages (Section~\ref{sec:loops}),
\item \emph{putback} messages that implement lens functionality,
\item those messages that are listed as syntactic sugar in Table~\ref{tab:syntacticsugar},
\item messages specific to the type. 
\end{itemize}

As a convention, all messages that are normally only used via some sort of syntactic sugar end with an underscore in their name.

\subsection{Collections and Collectors}
The built-in types that can be used as collectors and collections are: List, Vector, Set, Map and String. In the case of maps,
the elements of the collection are pairs (vectors of length 2) where the first element of the pair represents the key, and the second element the value the key is mapped to. For strings, the elements of the collection are strings of length 1, consisting of a single Unicode code point. 

All built-in collection/collector types implement the messages described in Table~\ref{tab:generalcollectmessages}.
\begin{table}
\caption{Additional messages for collection/collector $c$}
\begin{tabular}{c|l}
\textbf{Message} & \textbf{Description} \\\hline
\babelsrc{$c$.isEmpty} & \tabparbox{checks if $c$ is an empty collection} \\\hline
\babelsrc{$c$.empty} & \tabparbox{an empty collection that has the same type as $c$} \\\hline
\babelsrc{$c$.size} & \tabparbox{the size of collection $c$} \\\hline
\babelsrc{$c + x$} & \tabparbox{adds $x$ as member to collection $c$} \\\hline
\babelsrc{$c\ \text{++}\ d$} & \tabparbox{adds all members of $d$ as members to $c$} \\\hline
$c$ \verb!-! $x$ & \tabparbox{removes from $c$ all members that are equal to $x$} \\\hline
$c$ \verb!--! $d$ & \tabparbox{removes from $c$ all members that are equal to an element in $d$} \\\hline
$c$ \verb!**! $d$ & \tabparbox{removes from $c$ all members that are not equal to an element in $d$} \\\hline
\babelsrc{$c$.head} & \tabparbox{the first element of $c$} \\\hline
\babelsrc{$c$.tail} & \tabparbox{all elements of $c$ except the first one} \\\hline
\babelsrc{$c$.atIndex $\ i$} & \tabparbox{the $i$-th element of $c$} \\\hline
\babelsrc{$c$.indexOf $\ x$} &\tabparbox{ the lowest $i$ such that \babelsrc{$c$.atIndex $\ i\ ==\ x$} } \\\hline
\babelsrc{$c$.contains $\ x$} & \tabparbox{checks if $c$ contains $x$} \\\hline
\babelsrc{$c$.take $\ n$} & \tabparbox{forms a collection out of the first $n$ elements of $c$} \\\hline
\babelsrc{$c$.drop $\ n$} &\tabparbox{ forms a collection by dropping the first $n$ elements of $c$} \\\hline
$c / f$  & \tabparbox{maps the function $f$ over all elements of $c$} \\\hline
$(c * f)\ a_0$ & \tabparbox{folds $f$ over $c$ via $a_{i+1}$ = $f\ (c_i,\ a_i)$} \\\hline
$c\ \text{\textasciicircum}\ f$  & \tabparbox{filters $c$ by boolean-valued function $f$} \\\hline
$c$ \verb+//+ $f$  & \tabparbox{map created by all key/value pairs $(x,\ f(x))$ where $x$ runs through the elements of $c$; later pairs overwrite earlier pairs} 
\end{tabular}
\label{tab:generalcollectmessages}
\end{table}

\subsection{Type Conversions}
Many built-in types have conversions defined between them. Some of these type conversions are automatic, and some are not. Table~\ref{tab:typeconversions} lists all of them. Note that a "yes" and "auto" do not necessarily mean that \emph{all} values of the source type can be converted to the destination type. For example, \verb+5^1000+ cannot be converted to \texttt{real}, and \texttt{5.1} cannot be automatically converted to \texttt{int}, but nevertheless \texttt{5.1 :> int == 5} holds.
\begin{table}
\caption{Type Conversions \texttt{src :> dest}}
\begin{tabular}{cc|cccccccc}
   & \textbf{dest}& \texttt{int} & \texttt{bool} & \texttt{real} & \texttt{string} & \texttt{list} & \texttt{vect} & \texttt{set} & \texttt{map}  \\
 \textbf{src} & \\\hline
\texttt{int} & & - & yes & auto & yes & no & no & no & no \\
\texttt{bool} & &  yes &  - & no  & yes & no & no & no & no\\
\texttt{real} & & auto & no & - & yes & no & no & no & no \\
\texttt{string} & & yes & yes & yes & -  & yes & yes & yes & no\\
\texttt{list} & & no & no & no & no & - & auto & yes & yes\\
\texttt{vect} & & no & no & no & no & auto & - & yes & yes\\
\texttt{set} & & no & no & no & no & yes & yes & - & yes\\
\texttt{map} & & no & no & no & no & yes & yes & yes & -\\
\end{tabular}

\label{tab:typeconversions}
\end{table}

In the rest of this section we enumerate for each built-in type all messages that are supported by this type. Collector and collection messages are excluded from this enumeration.

\subsection{Integer}
Integers can be arbitrarily large and support the usual operations (\texttt{+}, binary and unary \texttt{-}, \texttt{*}, \texttt{\^}, \babelsrc{div}, \babelsrc{mod}). Division and modulo are Euclidean.

The expression $a$ \texttt{to} $b$ denotes the list of values $a$, $a+1$, $\ldots$, $b$. The expression $a$  \texttt{downto} $b$ denotes the list of values  $a$, $a-1$, $\ldots$, $b$. 

\subsection{Reals}
Reals are intervals bounded by floating point numbers. For more information on them see Section~\ref{sec:reals}.

\subsection{String} Implements no messages specific for strings. The semantics of the \babelsrc{indexOf} and \babelsrc{contains} messages is extended with respect to the usual collection semantics to not only search for strings of length 1, but arbitrarily long strings.

\subsection{List and Vector} Those messages specific to lists and vectors are listed in Table~\ref{tab:listvectorops}.  
\begin{table}
\caption{List/Vector-specific messages}
\begin{tabular}{c|c}
$l$ $i$ & \tabparbox{same as \babelsrc{$l$.atIndex $\ i$}}\\\hline
$-l$ & \tabparbox{reverses $l$}
\end{tabular}
\label{tab:listvectorops}
\end{table}

\subsection{Set}
There are no messages specific to sets except that for a set $s$ the expression $s\ x$ is equivalent to 
$s.\babelsrc{contains}\ x$ and therefore tests if $x$ is an element of $s$. 

\subsection{Map}The messages specific to maps are listed in Table~\ref{tab:mapops}. Note that the operators \texttt{-}, \texttt{--}, \texttt{**} and \texttt{//} deviate in their semantics from the usual collection semantics, but that the operators  \texttt{+} and \texttt{++} \emph{do} behave according to the usual collection semantics. 
\begin{table}
\caption{Map-specific Messages}
\begin{tabular}{c|c}
\babelsrc{$c$.contains $\ x$} & \tabparbox{checks if $c$ contains $x$} \\\hline
\babelsrc{$c$.containsKey $\ k$} & \tabparbox{checks if $c$ contains $(k, v)$ for some $v$} \\\hline
$m$ \verb!+! $(k, v)$ & \tabparbox{map created from the map $m$ by associating $k$ with $v$}\\\hline
$m$ \verb!-! $k$ & \tabparbox{map created from the map $m$ by removing the key $k$}\\\hline
$m$ \verb!++! $n$ &  \tabparbox{map created from the map $m$ by adding the key/value pairs that are elements of $n$}\\\hline
$m$ \verb!--! $n$ & \tabparbox{map created from $m$ by removing all keys that are elements of $n$}\\\hline
$m$ \verb!**! $n$ & \tabparbox{map created from $m$ by removing all keys that are not elements of $n$} \\\hline
$m$ $k$ & \tabparbox{returns the value $v$ associated with $k$ in $m$, or returns a dynamic exception with parameter \texttt{DomainError} if no such value exists} \\\hline
$m$ \verb!//! $f$ & \tabparbox{map created by applying the function $f$ to the key/value pairs $(k,\ v)$ of $m$, yielding key/value pairs 
$(k, \ f(k,\ v))$}
\end{tabular}
\label{tab:mapops}
\end{table}

\subsection{Object}
Custom objects respond exactly to the set of messages defined in their body, with one exception: if the custom object implements all four of these messages:
\begin{itemize}
\item \babelsrc{collector_add_}
\item \babelsrc{collector_close_}
\item \babelsrc{empty}
\item \babelsrc{collector_iterate_}
\end{itemize}
then the custom object automatically inherits standard implementations for all messages listed in Table~\ref{tab:generalcollectmessages} and for the three type conversions to \texttt{list},\texttt{vector} and \texttt{set}. These standard implementations are shadowed by actual implementations the custom object might provide.

\section{What's Next?}

Where will Babel-17 go from here? There are two dimensions along which Babel-17 has to grow in order to find adoption.

\vspace{0.2cm}
\noindent\textbf{Maturity and breadth of implementations}
While there will probably always be a reference implementation of Babel-17 that strives for simplicity and clarity and sacrifices performance to achieve this, Babel-17 will need quality implementations on major computing platforms like JavaVM, Android, iOS and HTML 5. A lot of the infrastructure of these implementations can be shared, so once there is a speedy implementation of Babel-17 running for example on the Java Virtual Machine, the other implementations should follow more quickly. For example, there are many many opportunities for both static and particularly dynamic optimizations of the standard library implementation.

\vspace{0.2cm}
\noindent\textbf{Development of the language} 
Babel-17 is a dynamically typed purely functional programming language with a module system that has powerful encapsulation mechanisms.
Some way of dealing with state, user interfaces, and communication with the outside world will be added to future versions of Babel-17. 

\bibliographystyle{abbrvnat}


\end{document}